# *Beyond Predicted zT: Machine Learning Strategies for the Experimental Discovery of Thermoelectric Materials*

Shoeb ATHAR and Philippe JUND*

ICGM, Univ Montpellier, CNRS, ENSCM, 34293 Montpellier, France

*Corresponding author. E-mail address: philippe.jund@umontpellier.fr (Philippe JUND)

## Abstract

The discovery of high-performance thermoelectric (TE) materials for advancing green energy harvesting from waste heat is an urgent need in the context of looming energy crisis and climate change. The rapid advancement of machine learning (ML) has accelerated the design of thermoelectric (TE) materials, yet a persistent "gap" remains between high-accuracy computational predictions and their successful experimental validation. While ML models frequently report impressive test scores ($R^2$ values of 0.90–0.98) for complex TE properties ($zT$, power factor, and electrical/thermal conductivity), only a handful of these predictions have culminated in the experimental discovery of new high-$zT$ materials. In this review, we identify and discuss that the primary obstacles are poor model generalizability—stemming from the "small-data" problem, sampling biases in cross-validation, and inadequate structural representation—alongside the critical challenge of thermodynamic phase stability. Moreover, we argue that standard randomized validation often overestimates model performance by ignoring "hidden hierarchies" and clustering within chemical families. Finally, to bridge this gap between ML-predictions and experimental realization, we advocate for advanced validation strategies like PCA-based sampling and a synergetic active learning loop that integrates ML

"fast filters" for stability (e.g., GNoME) with high-throughput combinatorial thin-film synthesis to rapidly map stable, high-*zT* compositional spaces.

## 1. Introduction

The urgency to find green energy solutions and improve energy efficiency, driven by fossil fuel depletion and increased global consumption, highlights the need for advanced energy recovery, particularly since more than 60% of fossil energy is lost as waste heat [1]. Highly efficient thermoelectric (TE) materials, which convert this heat directly into electricity using the Seebeck effect, are therefore highly relevant. Their performance is measured by the figure of merit $\boldsymbol{zT} = \frac{\boldsymbol{S^2\sigma}}{\boldsymbol{\kappa}}\boldsymbol{T}$ (where S is the Seebeck coefficient ($VK^{-1}$), T the working temperature (K), σ the electrical conductivity ($Sm^{-1}$), and $\boldsymbol{\kappa}$ the thermal conductivity ($Wm^{-1}K^{-1}$)), where maximizing the power factor ($\boldsymbol{S^2\sigma}$) and minimizing thermal conductivity ($\boldsymbol{\kappa}$) is the goal [2]. Despite significant research progress yielding high-performance materials, their large-scale application is limited by issues such as the cost and toxicity of constituent elements (e.g., Pb, Te, Ge) and poor mechanical stability under operational loads [3, 4]. Consequently, the search for efficient thermoelectric materials composed of affordable, non-toxic, and earth-abundant elements with superior mechanical properties remains a primary research priority.

The unabated progress in artificial intelligence has opened new horizons into the accelerated discovery and design of new functional materials [5]. Machine Learning (ML) being a central technology of material informatics has emerged as a powerful technique to design and screen materials with required functionalities. [6, 7]. From lithium-ion batteries [8] to perovskite materials [9], ML has proved to be an effective tool in advancing materials discovery. Thermoelectric (TE) materials, being no exception, have also benefitted from ML assisted compositional design [10] in the last two decades evidenced by a surge in number of papers published on the topic [11]. Several studies have demonstrated that ML can not only be used to predict individual transport properties such as Seebeck coefficient (S), electrical (σ) or thermal conductivity (κ), but also complex properties like power factor (PF) and $zT$.

Table 1 chronologically summarizes the notable works on ML application for TE materials discovery [12-36]. Impressive $R^2$ values of ~0.98 for σ [18], ~0.95 for PF [19], >0.93 for κ [16], ~0.90 for $zT$ [32], have been reported as test scores for the ML models trained on experimental datasets. Despite these remarkable metrics, there are only a handful of works reporting experimental validations of promising TE materials predicted through high-throughput screening (HTS) – a globally consistent trend for TE materials [11]. For experimental validation on $zT$, there are four noteworthy examples where the ML-predictions culminated in experimental discovery of a high $zT$ TE material. Jia et al. (2022) [25] demonstrated ML-assisted experimental discovery of *p-type* $Sc_{0.7}Y_{0.3}NiSb_{0.97}Sn_{0.03}$ and *n-type* $Sc_{0.65}Y_{0.3}Ti_{0.05}NiSb$ with maximum $zT$ values of ~0.5 at 925 K and ~0.3 at 778 K, respectively. However, these values are still way inferior compared to the state-of-the-art half-Heusler (hH) materials, for example ~1.45 for $Nb_{0.88}Hf_{0.12}FeSb$ [37]. To the best of our knowledge, there is no other work on half-Heuslers in the literature demonstrating an experimental validation of ML-predictions for a higher $zT$. Certainly, following this study, there have been discoveries of hH materials with better $zT$s without using the data-driven route [38]. Another example is for SnSe-doped materials where Lee et al. (2022) [26] used experimentally measured TE properties of 263 samples of doped SnSe to develop Gradient-Boosted Regression Trees (GBRTs) for predicting and screening SnSe-based compounds. The model was trained using a feature vector comprising of compositional information, elemental properties and high-throughput DFT-generated electronic structures of supercell models of all possible dopants. A new Y-doped SnSe compound exhibiting a very high *zT* above 2.0 was screened and experimentally validated. Interestingly, Y-doped compositions were already present in dataset. In fact, out of the top five dopants (Ge, Pb, Y, Cd, and As) predicted in this study, except As, the other four were also present. Though the extrapolation ability of model was demonstrated for V-doping, the experimentally achieved *zT* values (<1) were significantly lower than those of these four compositions.

Zhong et al. (2023) [29] employed SISSO in an active learning (AL) framework to identify a trend in the materials $Cu_{1-x}Ag_xGaTe_2$ for high *zT*, through physically informed descriptors. Using a small dataset of ~600 experimental *zT*s at different compositions and temperature for ternary chalcogenides and only six elemental features, they predicted as well as experimentally synthesized several high-performing TE chalcogenides. Finally, the *p-type* $Cu_{0.45}Ag_{0.55}GaTe_2$ was reported to have a very high experimental figure of merit (*zT* ~1.90 at 770 K). Unfortunately, the presence of Te in the best screened candidate is not helpful. Recently, the work by Long et al. (2025) [36] was a seminal example on application of Generative adversarial networks (GAN) for TE materials discovery leading to the experimental synthesis and validation of the $Mg_{3.1}Sb_{0.5}Bi_{1.497}Te_{0.003}$ with a peak *zT* of ~0.75 at 300 K. However, analysis of their training data revealed at least 92 compounds in the same Mg-Sb-Bi-Te quaternary system, including those very close to the 'discovered' material, such as $Mg_{3.55}Bi_{1.490}Sb_{0.480}Te_{0.03}$, $Mg_{3.55}Bi_{1.470}Sb_{0.500}Te_{0.03}$, etc. This, therefore, suggests that the model was operating within a heavily sampled 'safe zone.' Consequently, the 'experimental validation' reflects the model's ability to interpolate within a known high-*zT* chemical space rather than its ability to navigate toward unexplored spaces. Their results, therefore, serve more as a proof of stoichiometric optimization than true material discovery.

Table 1: Chronological summary of notable ML works on TE materials (non-exhaustive)

| **Reference** | **Target property (TP)** | **Database (type)** | **Size** | **Structural prototype** | **(Best) model** | **(Best) test Scores** | **DFT valid.** | **Exp. valid.** | **Exp. realized TP** |
|---|---|---|---|---|---|---|---|---|---|
| **Carrete et al. (2014) [12]** | $\kappa_{lat}$ | In-lab (DFT) | 32 materials | Half-Heusler | RF | Spearman rank correlation coeff. ~ 0.74 | Yes | No | |
| **Seko et al. (2015) [13]** | $\kappa_{lat}$ | In-lab (DFT) | 101 materials | rocksalt, zinc-blende and wurtzite-type | Bayesian Optimization | Not given | Yes | No | |
| **Furmanchuk et al. (2017) [14]** | S | MRL + Literature (exp.) | 130 materials | Non-specific | RF | $R^2 \sim 0.88$ | No | No | |
| **Hou et al. (2019) [15]** | PF | In-lab (exp.) | Not given | $Al_{23.5+x}Fe_{36.5}Si_{40-x}$ compounds | GPR | $R^2 \sim 0.99$ | Yes | **Yes** | **PF ~ 670 μW $m^{-1}K^{-2}$ at 510 K** |
| **Chen et al. (2019) [16]** | $\kappa_{lat}$ | Literature (exp.) | 100 materials | Non-specific | GPR | RMSE ~ 0.28 $Wm^{-1}K^{-1}$ (log-scaled); $R^2 \geq 0.93$ | No | No | |

Contd…

| Reference | Target property (TP) | Database (type) | Size | Structural prototype | (Best) model | (Best) test scores | DFT valid. | Exp. valid. | Exp. realized TP |
|---|---|---|---|---|---|---|---|---|---|
| **Juneja et al. (2019) [17]** | $\kappa_{lat}$ | In-lab (DFT) | 120 materials | binary, ternary, quaternary compounds | GPR | RMSE ~0.21 $Wm^{-1}K^{-1}$ (log-scaled); $R^2$~ 0.99 | No | No | |
| **Mukherjee et al. (2020) [18]** | σ | Literature (exp.) | 124 materials | binary, ternary, quaternary compounds | GBR | RMSE ~0.22 $Scm^{-1}$ (log-scaled); $R^2$~ 0.98 | No | No | |
| **Sheng et al. (2020) [19]** | PF | In-lab (exp.) | 158 materials | diamond-like pnictides, chalcogen-ides | GBR | $R^2$~ 0.95 | Yes (AL) | No | |
| **Wang et al. (2020) [20]** | $\kappa_{lat}$ | AFLOW ( AGL method) | 5486 materials | Non-specific | XGBoost | RMSE/MAE ~0.36/0.259 $Wm^{-1}K^{-1}$ (log-scaled); $R^2$ ~0.90 | Yes | No | |
| **Gan et al. (2021) [21]** | *zT* | In-lab (DFT) | 70 materials | Layered IV-V-VI semicond-uctors | NN | MSE ~0.008 ; $R^2$ ~0.952 | Yes | No | |

Contd…

| Reference | Target property (TP) | Database (type) | Size | Structural prototype | (Best) model | (Best) test scores | DFT valid. | Exp. valid. | Exp. realized TP |
|---|---|---|---|---|---|---|---|---|---|
| **Na et al. (2021) [22]** | *zT* | MRL (exp.) | 573 data points | Non-specific | NN | MAE ~ 0.06; $R^2$ ~0.86; | No | No | |
| **Yuan et al. (2022) [23]** | S | In-lab + literature+ Materials Project (DFT) | 122 Half-/129 full-Heusler materials | Half-/full-Heuslers | NN | RMSE ~ 39.4/31.4 $\mu VK^{-1}$; $R^2$ ~96% /98% for n-type/p-type | Yes | No | |
| **Bhattacharjee et al. (2022) [24]** | $\kappa_{lat}$ | Literature (DFT) | 110 materials | Half-Heusler | SISSO | $R^2$ ~0.97 | No | No | |
| **Jia et al. (2022) [25]** | *zT* | Materials Project (DFT) | 456 materials | Half-Heusler | Unsupervised learning (*sklearn.cluster)* | Not applicable | No | **Yes** | ***zT* ~ 0.5 at 925 K** |
| **Lee et al. (2022) [26]** | *zT* | In-lab (exp.) | 263 compositions | SnSe-based materials | GBRT | MAE ~ 0.102; $R^2$ ~0.756 | No | **Yes** | ***zT* ~2.0 at 798 K** |

Contd…

| Reference | Target property (TP) | Database (type) | Size | Structural prototype | (Best) model | (Best) test scores | DFT valid. | Exp. valid. | Exp. realized TP |
|---|---|---|---|---|---|---|---|---|---|
| **Plata et al. (2022) [27]** | $\kappa_{lat}$ | In-lab (DFT) | 20 materials | $ABX_2$ (I–III–$VI_2$) chalcopyrites | multi-linear regression (HiPhive package) | MAE $< 1.5$ $Wm^{-1}K^{-1}$ | Yes | No | |
| **Barua et al. (2023) [28]** | κ | In-lab (exp.) | 776 Compositions | SnSe-based materials | XGBoost | RMSE/MAE ~0.07/ 0.05 $Wm^{-1}K^{-1}$; $R^2$~0.84 | No | **Yes** | **κ ~ 0.80 $Wm^{-1}K^{-1}$ at 300 K** |
| **Zhong et al. (2023) [29]** | *zT* | In-lab (exp.) | 600 data points | A–B–$C_2$ ternary chalcogenides | SISSO | RMSE ~ 0.14 | No | **Yes (AL)** | ***zT* ~1.9 at 770 K** |
| **Borg et al. 2023 [30]** | *zT*, κ | Starrydata2 | 626 compositions | 111-type materials | RF | Not given | No | No | |
| **Barua et al. (2024) [31]** | κ | Starrydata2 + MRL + CHER + in-lab (exp) | ~200,000 data points | Non-specific | XGBoost | RMSE/MAE ~ 0.52/0.40 W $m^{-1}K^{-1}$; $R^2$~0.89 | No | No | |

Contd…

| Reference | Target property (TP) | Database (type) | Size | Structural prototype | (Best) model | (Best) test scores | DFT valid. | Exp. valid. | Exp. realized TP |
|---|---|---|---|---|---|---|---|---|---|
| **Jia et al. (2024) [32]** | *zT* | Starrydata 2 | 7,295 compositions | Non-specific | GBDT | $R^2$ ~0.90 | Yes | No | |
| **Parse et al. (2024) [33]** | *zT* | Starrydata 2 | 23,662 data points | Non-specific | XGBoost | MAE ~ 0.103; $R^2$~ ~ 0.815 | No | No | |
| **Barua et al. (2024) [34]** | *zT* | Starrydata 2 + MRL | 160,000 data points | Non-specific | XGBoost | RMSE/MAE~ 0.156/0.091; $R^2$~0.80 | No | No | |
| **Posligua et al 2025 [35]** | *zT* | Literature + Starrydata 2 | 4000 compositions | Skutterudites | NN | RMSE~0.128; $R^2$~0.863 | No | No | |
| **Long et al. 2025 [36]** | *zT* | MRL+ Literature | 3083 Compositions | Non-specific | GAN | MSE = 0.032 | Yes | **Yes** | ***zT* ~0.75 at 300 K** |

**Acronyms:** GBDT = Gradient Boosting Decision Tree; GPR = Gaussian process regression; SISSO = Sure Independent Screening – Sparsifying Operator; RF = Random Forest; NN = Neural Networks; XGBoost = Extreme Gradient Boosting; GBR = Gradient Boost Regression; GAN = Generative adversarial networks

The limited success in ML-driven experimental discovery of TE materials demonstrates a clear gap between ML-predictions and their realization through experimental synthesis and characterization. In this review, we argue that this gap originates from both the poor generalizability of published models as well as phase stability of predicted materials. Therefore, in the following sections, we discuss the underlying reasons for this existing gap and propose the possible strategies to bridge it.

## 2. Poor model generalizability

The poor generalizability of ML models for predicting the TE properties of thermoelectric materials originates from several interconnected challenges primarily related to the nature of the available data, sampling bias in data-preprocessing, and the structural complexity of the materials.

### a. The small-data problem

Machine Learning requires data and several research groups have attempted to build extensive datasets of TE materials containing the different TE properties. Table 2 depicts the list of publicly available datasets along with the numbers of compositions -data collated from [39-41] and original works [16, 22, 24, 31, 42-53] - that can be utilized for ML studies of TE materials. These datasets are either based on theoretical calculations, such as DFT, or from experimental data generated *in-house* or reported in the literature. Among the DFT-based datasets, the one developed by Ricci et al. (2017) [44] is the largest dataset containing the Seebeck coefficient and electrical conductivity values of approximately 48000 compositions at different temperatures calculated with the BoltzTrap code [54]. The JARVIS-DFT database [47] also has an impressive record of BoltzTrap-calculated S, σ, and PF values of about 21900 compositions. While these large databases can be readily used for ML, an exhaustive TE characterization of a material warrants the inclusion of all TE properties making up the TE figure-of-merit $zT$. Nevertheless, the absence of thermal conductivity values in these datasets, (particularly lattice thermal conductivity) renders this task infeasible. This may be attributable to the relatively higher cost of DFT calculations of $\kappa_{lat}$ against electronic properties. This cost becomes further prohibitive when dealing with doped compounds since the supercells needed to calculate the phonon spectrum become huge and not easy to handle in *ab initio* simulations. The other DFT-based datasets, by Jaafreh et al (2021) [48], Tranas et al. (2021) [49], and Bhattacharjee et al. (2022) [24], containing $\kappa_{lat}$ entries lack the electronic part of $zT$ and are much smaller.

Table 2: The list of publicly available datasets that can be utilized for ML of TE materials.

| **Database** | **Source** | **No. of Compositions** | **Properties** | **Year** | **Reference** |
|---|---|---|---|---|---|
| Wang et al. | Theory | 2,585 | PF | 2011 | [42] |
| MRL (UCSB) | Experiment | 524 | $zT$, S, σ, κ | 2013 | [43] |
| Ricci et al. | Theory | 48,000 | S, σ, $\kappa_{el}$ | 2017 | [44] |
| Xi et al. | Theory | 161 | PF | 2018 | [45] |
| Starrydata2 | Experiment | ~50,000 | $zT$, S, σ, κ | 2019 | [46] |
| Chen et al. | Experiment | 100 | $\kappa_{lat}$ | 2019 | [16] |
| JARVIS-DFT | Theory | 21,900 | S, σ, PF | 2020 | [47] |
| Jaafreh et al. | Theory | 119 | $\kappa_{lat}$ | 2021 | [48] |
| Tranås et al. | Theory | 122 | $\kappa_{lat}$ | 2021 | [49] |
| MIP-3d | Theory | 4,400 | S, σ | 2021 | [50] |
| Bhattacharjee et al. | Theory | 110 | $\kappa_{lat}$ | 2022 | [24] |
| CHER | Experiment | 328 | $zT$, S, σ, κ | 2022 | [51] |
| ESTM | Experiment | 880 | $zT$, S, σ, κ | 2022 | [22] |
| ChemDataExtractor | Experiment | 10,641 | $zT$, S, σ, κ | 2022 | [52] |
| UWAT_TE | Experiment | 150 | $zT$, S, σ, κ | 2024 | [31] |
| GPTArticleExtractor | Experiment | 7,123 | $zT$, S, σ, κ | 2025 | [53] |

An exhaustive high-throughput screening of a material class, half-Heusler or otherwise, with all possible dopants and concentrations, therefore, requires an experimental dataset containing all TE properties of experimentally synthesized pure and doped materials. Gaultois et al. (2013) [43] were the first to publish a manually extracted (plotdigitizer) [55] experimental TE dataset (of *currently* 524 materials) containing all TE properties: $zT$, S, $\sigma$, $\kappa$. With the advent of Large Language Models based data-mining and data-curation, larger experimental TE datasets, such as Starrydata2 [46, 56], ChemDataExtractor [52], GPTArticleExtractor [53], have evolved lately. Among the experimental datasets, to the best of our knowledge, Starrydata2 is the largest publicly available TE materials dataset hosting the TE properties of ~50,000 samples [41] obtained under different temperatures from 10,074 publications relevant to various TE systems (e.g. Chalcogenides, Zintl, Silicides, half-Heusler compounds) [46, 56].

These experimental datasets are hierarchical in nature. As shown in Figure 1, there are three levels of hierarchies. At the top, the "material" represents the general stoichiometry of the compound of interest for any range of concentrations. Then "compositions" represent the exact stoichiometry of the compound with explicit concentration values. Lastly for a given composition there are several temperature points for the TE property under consideration (e.g.

*zT*). Therefore, while there can be tens of thousands of datapoints present in the dataset, it is important to consider the top hierarchy “materials” for measuring the data volume. For instance, in Starrydata2, while the number of datapoints is >9000 for half-Heusler TE materials, the number of compositions and materials represented by these datapoints are only 367 and 132, respectively [57]. To qualify as big data, these datasets must not only be large in data volume but also be high in data quality as well as rich in data diversity [58]. Unfortunately, these datasets incorporate a significant degree of noise and inconsistent, and in-fact inaccurate data, originating from multi-source experimental data for same composition ambiguous nomenclature in the literature, or limitations of LLM-assisted data-curation [32]. As for the data diversity, with the example of hH structural prototype, in our recent work [57], we have compared the combinatorial chemical space for predicted hH systems from Materials Project (MP) [59] and unexplored hH systems (with single or double doped sites) to the existing hH TE materials experimentally reported in the literature. We mapped the chemical space of half-Heusler materials by representing each $A_{1-x_{A'}} A'_{x_{A'}} B_{1-x_{B'}} B'_{1-x_{B'}} C_{1-x_{C'}} C'_{x_{C'}}$ hH material as a 114-dimensional vector, derived from 19 physical properties per element site. By applying Uniform Manifold Approximation and Projection (UMAP) [60] dimensionality reduction, we projected this high-dimensional data into a visual map to compare millions of unexplored combinatorial permutations against known stable systems from the experimental literature and the Materials Project database (those with a hull distance < 0.05 eV/atom). Further details of the calculation can be found in the original paper [57]. The resulting UMAP projection, shown in Figure 2, reveals that despite the substantial volume of available data points, the half-Heusler materials present in the literature exhibit remarkably low chemical diversity. Even the integration of the Materials Project (MP) database fails to significantly bridge this gap, demonstrating a discrepancy of several orders of magnitude between known materials and the ~23.6 million potential ‘materials’ within the hH family. While this analysis focused on the hH structural prototype as a representative case study, the results are indicative of a broader systemic challenge. Given the combinatorial explosion inherent in other prominent thermoelectric families—such as skutterudites and complex chalcogenide solid solutions—this data sparsity is likely a universal feature for TE materials. Although quantifying the exact overlap across all material classes is currently computationally intensive due to the high-dimensional (114-D) feature representation required (in our method), the hH system serves as a proxy for global unexplored regions. These findings underscore the need for future high-performance computing efforts to quantify these overlap rates and guide more targeted experimental exploration across the entire gamut of TE material families.

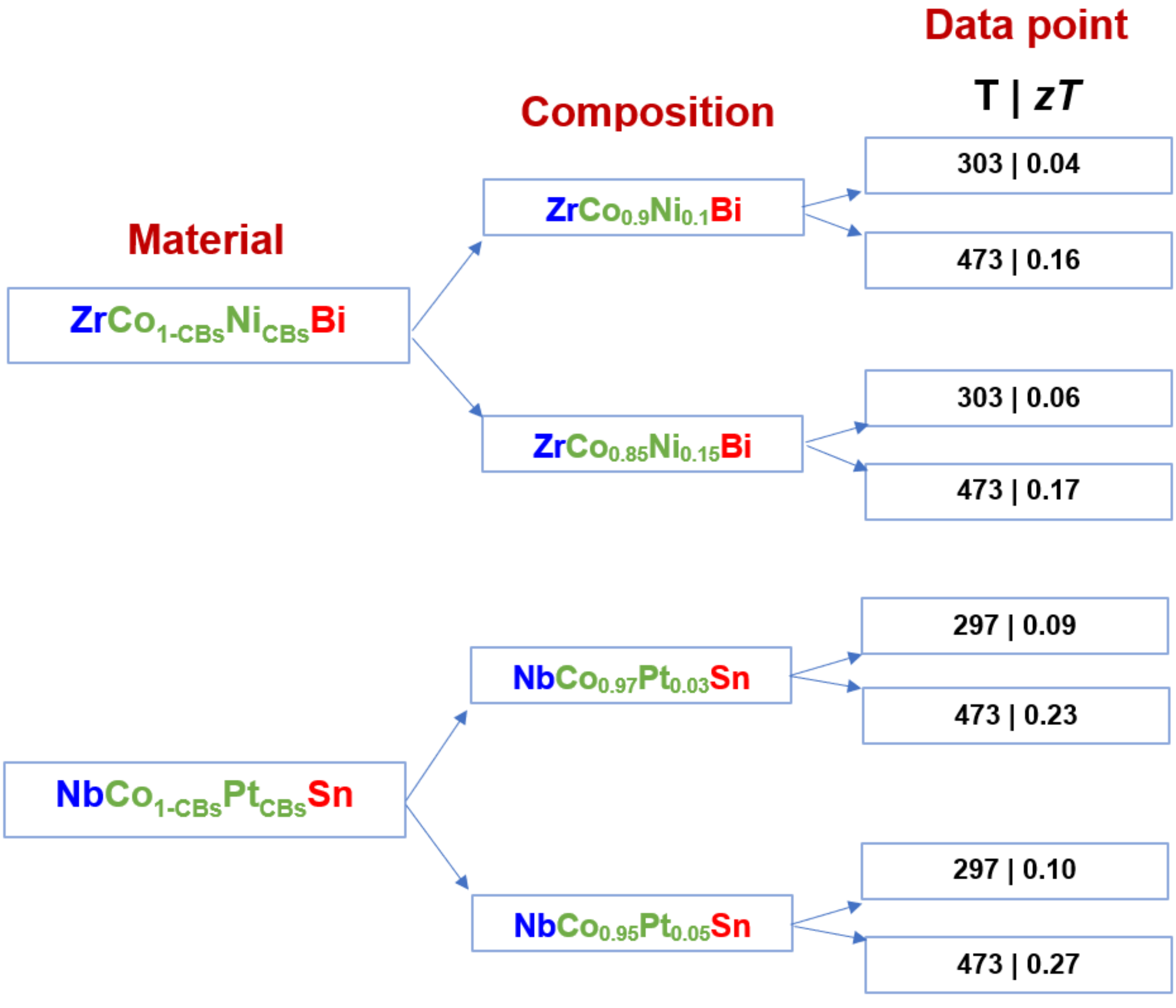

Figure 1: Hierarchical structure of an experimental thermoelectric materials dataset.

Our previous work [57] also proposes a rigorous bin-filtering based data curation strategy to systematically address the problem of data quality in these “large” datasets. This method is a statistical data curation strategy based on "round-robin errors" (typical ±15 % measurement uncertainties [61]) to mitigate inaccuracies in large-scale experimental datasets. It operates by collecting all reported curves for a given composition and grouping their maximum figure-of-merit ($zT_{max}$) values into bins defined by the 15% (or any other user-defined) error threshold. By selecting the most populated bin, the method identifies the consensus property range, effectively isolating and removing outliers caused by mislabeling or non-representative synthesis routes. A high-quality reference curve is then selected based on maximum "information gain" (e.g., number of doped variants reported), ensuring the resulting dataset is both physically consistent and optimized for machine learning reliability.

However, the issue of data volume and diversity requires multi-institutional collaborative efforts to create a chemically comprehensive and curated dataset covering all structural prototypes of TE materials. The feasibility of Active Learning (AL) is well-established for effectively overcoming data limitations in materials discovery in general [62, 63] and has been specifically demonstrated for thermoelectric (TE) materials by Zhong et al. (2023) [29]. Notably, this is the only study in our survey that successfully resulted in the discovery of a 'new material' through an AL-driven approach. By iteratively selecting only the most critical materials for the desired TE properties, this strategy can minimize the experimental workload while effectively enriching the dataset and improving prediction accuracy. The intrinsic data scarcity in TE materials can further be overcome by guiding future experimental efforts based on the least explored parts of the chemical space. Such targeted experiments can generate high-quality, compositionally diverse data and pave the way for a truly ‘big-data’-based ML for TE materials discovery.

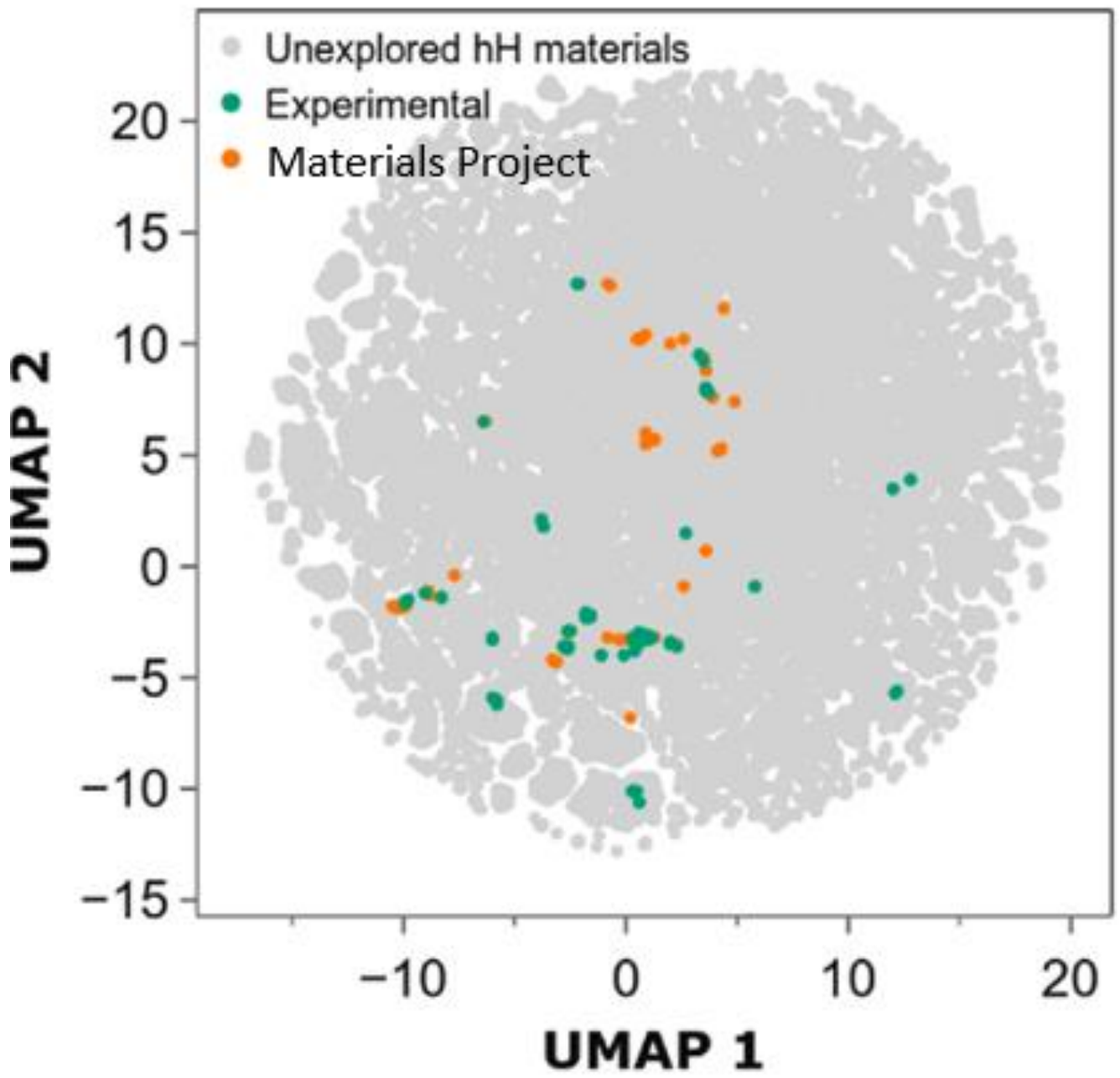

Figure 2: Unexplored half-Heusler materials (grey) compared to experimental thermoelectric hHs till date (green) and predicted hHs structures from Materials Project (MP) database (orange) [**57**]

### b. The sampling bias

In machine learning, a theoretical guarantee of generalization is provided by Hoeffding's inequality [64, 65] which is a statistical bound on the probability that a model's empirical error (in-sample error) is a good proxy for its true error (error on unseen data/out-sample error). The Hoeffding's inequality (for a perceptron model) is given as:

$$P\,[|E_{in}(h) - E_{out}(h)| < \epsilon] = 2e^{-2\epsilon^2 N} \text{ for any } \epsilon > 0 \qquad \text{Equation 1}$$

Where P [E] denotes the probability of an event E, $E_{in}$ (h) and $E_{out}$(h) are the in-sample/training error and the out-sample error for a given hypothesis/model 'h', $\epsilon$ is a threshold of the statistical bound, N is the number of training data points

This inequality mathematically means that the probability of these two errors diverging by more than a certain threshold ($\epsilon$) is equal or smaller than $2e^{-2\epsilon^2 N}$. This implies that as the number of data points (N) increases, the probability of the in-sample error ($E_{in}$ (h)) to be a bad estimate for the out-of-sample error decreases. The Hoeffding's inequality is not only crucial for understanding how good a learned model will behave on new data for a given sample size but also provides a statistical guarantee for the generalizability of any machine learning model [65].

However, for a hypothesis set $H = \{h_1, h_2, h_3, ..., h_m\}$, there are M number of possible hypotheses, and the learning algorithm can pick any hypothesis as a final model 'g' based on the data. Therefore, "$|E_{in}(g) - E_{out}(g)| < \epsilon$" should be bound in a way that it becomes independent of whichever 'g' the learning algorithm selects. Since, g is one of the $h_M$'s regardless of the algorithm and sample, following the union bound rule we get [65]:

"$P[|E_{in}(g) - E_{out}(g)| < \epsilon] \leq P[|E_{in}(h_1) - E_{out}(h_1)| < \epsilon]$ Equation 2

Or $P[E_{in}(h_2) - E_{out}(h_2)| < \epsilon]$

Or $P[E_{in}(h_3) - E_{out}(h_3)| < \epsilon]$

…

Or $P[E_{in}(h_m) - E_{out}(h_m)| < \epsilon]$

Therefore,

$$P[|E_{in}(g) - E_{out}(g)| < \epsilon] \leq \sum_{m=1}^{M} P[|E_{in}(h_m) - E_{out}(h_m)| < \epsilon] \quad \text{Equation 3}$$

Applying Hoeffding's inequality for M possible hypotheses gives us:

$$P[|E_{in}(g) - E_{out}(g)| < \epsilon] = 2Me^{-2\epsilon^2 N} \text{ for any } \epsilon > 0 \quad \text{Equation 4}$$

While this equation allows the learning algorithm to choose any hypothesis (from '$H$') based on $E_{in}$, while expecting it to gauge $E_{out}$, a disadvantage is that the probability $2Me^{-2\epsilon^2 N}$ is now

M times larger than that for a single hypothesis ($2e^{-2\epsilon^2 N}$). This is where the idea of a test set becomes helpful. Since 'g' is obtained through training on a different dataset; sampling $E_{in}$ (g) on the test set removes the limitation imposed by M choices of hypotheses such that we can recover the same theoretical guarantee that we had as if there were only a single hypothesis [65].

$$P\,[|E_{in}(g) - E_{out}(g)| < \epsilon] = 2e^{-2\epsilon^2 N} \text{ for any } \epsilon > 0 \qquad \text{Equation 5}$$

The theoretical guarantees of generalization, we have developed through equations 1 to 5, require that training and test distribution must be the same i.e. they must be generated from the same distribution of the input space. According to Abu-Mostafa et al. 2012 [65]: "If the data is (inadvertently) sampled in a biased way, learning will produce a similarly biased outcome." If the training data is generated with exclusion of a certain part of the input space, or chemical space, the final model trained with it may not generalize. Conversely, the test data obtained from only a small part of the chemical space, may not suffice as a true proxy to gauge the out-of-sample error. Therefore, a "fair" split of the train/test data, based on the chemical space of the dataset, is extremely crucial. Here, it is important to note the hierarchical structure of the experimental TE datasets as discussed previously. A train/test split based on the lower levels of the hierarchy (composition or temperature dependencies), can give misleading test scores as they would only reflect the interpolation ability of the model with respect to concentrations or measurement temperatures. While researchers are cautious in splitting the train/test subsets based on the top hierarchy of the materials [22, 32], i.e. group-k-fold cross-validation (CV), to the best of our knowledge, in the majority of ML works on TE materials, except randomization and k-fold cross validations, no consideration is given to fairly split the dataset based on the equivalent representation of chemical space.

TE materials datasets have underlying clustering due to the material's crystal structure (half-Heulsers, silicides, skutteridites, etc.) or chemical families within a given structural prototype, such as $A_{IV}NiSn$ ($A_{IV}$ = Ti, Zr, and Hf), $A_{V}FeSb$ ($A_{V}$ = V, Nb, and Ta), $A_{IV}CoSb$ ($A_{IV}$ = Ti, Zr, and Hf), etc. within half-Heuslers. These clusters add additional levels of "hidden hierarchy" in the dataset. In a complex, high-dimensional input space (for TE materials), a randomized test/train split might, purely by chance, can place most of the samples covering one critical axis of variance (e.g. all $A_{IV}NiSn$ or half-Heusler samples) into the training set, leaving the model untested on that axis. Even the standard randomized k-fold CV will be insufficient because it will inevitably split these structurally, chemically, or experimentally related clusters and treat every data point (or materials) independently. Since, in a clustered dataset, samples within the same cluster often share similar features, upon shuffling the data (materials) randomly, regardless of the degree of randomization or number of folds, points from the same cluster will likely end up in both the training and test sets simultaneously. The model can, therefore, "memorize" the specific characteristics of that cluster during training and then "recognizes" them in the test set. This leads to optimistic overestimation (excellent RMSE or $R^2$ values) of the model's performance, that will not generalize to entirely new, unseen clusters in the out-of-sample data. Therefore, while randomization and k-fold CV provide statistical fairness, ignoring the structure of the input space (the features) means poor representation of subgroups.

This issue can be addressed by clustering-based CV using, for example, agglomerative clustering with the ward linkage criterion as implemented in the "scipy" python library [66]. Such clustering can identify groups of TE materials with similar properties on the basis of their linkage distances. Another approach could be using dimensionality reduction technique, such as PCA, to sample the materials based on their position in the chemical space represented by their physical properties [49]. By analyzing the data variance, a PCA-based split can allow us to intentionally create folds that test the model's ability to extrapolate by sampling the training folds covering the dense, most common regions of the feature space (interpolation) and the test fold covering sparse, low-density regions, boundary conditions, or points far from the center of mass (extrapolation/outlier prediction). This will ensure a 'fair' estimation of model's generalizability through the RMSE or $R^2$ values. Tranas et al. (2021) [49] have reported impressive results on enhancing the ML prediction of the lattice thermal conductivity of half-Heuslers using active sampling with PCA. In our latest work [67], we have also proposed a rigorous PCA-based train/test splitting method for group-k-fold CV that guarantees an unbiased representation of the chemical space of the materials, and thereby, providing a more accurate measure of a model's generalizability for unseen materials. A hybrid approach may involve clustering the materials and then sampling representative materials from each cluster using PCA distances. These techniques can ensure that both train and test fold represent the actual dataset they were sampled from. However, if the dataset, on its own, does not accurately and adequately represent the true out-of-sample population due to the data sparsity, no amount of sophistication in the sampling technique can overcome the sampling-bias, thereby, leading to an optimistically biased performance estimate that will not generalize to the true, diverse population. This reminds us about the severity of "the small data" problem.

### c. Inadequate representation of structural diversity

TE materials are not only pristine single-crystals but are more often complex alloys, composites, or doped systems. Even with the advancements in ML techniques, discovering novel high-performance TE materials through an exhaustive search of compositional spaces is challenging due to the structural diversity and complexity of the TE materials containing several alloys and dopants [22]. While structure agnostic ML allows to predict the TE properties of all material classes using only the compositional information, the implicit assumption that two materials with the same chemical composition but different structures will have similar properties is problematic. Consequently, in high-throughput screening (HTS) of enormous unexplored compositional spaces, where any given composition can have many possible crystal structures, their generalizability is limited. An alternative is to include several structural and crystallographic features to train the model. However, using large feature spaces in combination with the small size of TE datasets, restricted to a handful of structural prototypes, may result in overfitting [10, 68]. Moreover, the inclusion of crystal structures in models may not be beneficial in HTS, as it is not possible to *a priori* generate structural information of unknown TE materials [22, 39, 69, 70]. A wiser approach is to focus on a specific class of materials by restricting the configurational space of materials [10]. In fact, all the ML models that could achieve experimental validation, listed in Table 1, were built on specific structural prototypes.

For the HTS, there is a rapidly growing field of "Composition-to-Structure" (C2S) ML models [71], such as CDVAE (Crystal Diffusion Variational Autoencoder) [72], Generative Adversarial Networks (GAN) for Crystal Structure Prediction [73], CrystalGAN [74], etc. which can predict the crystal structure from stochiometric compositions. These models are significantly faster than traditional methods like evolutionary algorithms or *ab initio* random structure searching [71]. They can be used to predict the structure of a new TE candidate from only its chemical formula; this output then serves as the input for structure-aware models or provides structural validation for structure-specific models. It should be noted that the C2S workflow is proposed here strictly as a strategy for HTS rather than for model training, which should remain grounded in high-fidelity experimental/DFT data or for specific structural prototypes. While this hierarchical approach introduces a risk of case-specific error propagation during the screening phase, this risk can be mitigated by using universal Machine Learning Interatomic Potentials (MLIPs), as proposed in section 3, to relax and validate C2S-predicted structures before property evaluation. This framework, therefore, can bypass the limitations of "structure-agnostic" models while still allowing for high-throughput screening.

## 3. Phase stability of predicted compositions

Addressing the generalizability problem of ML models alone cannot suffice for their culmination into successful experimental synthesis of promising TE materials. Predicting a thermoelectrically promising composition is the "easy" part whereas ensuring it occupies a stable (or usable metastable) pocket of the energy landscape is the harder physical constraint that renders most ML predictions experimentally invalid. A model might predict a chemical formula with a record-breaking *zT*, but if that composition is thermodynamically unstable, experimentalists waste time trying to synthesize "miracle materials" that simply never form or decompose into multiple phases, none of which have the desired properties. HTS is, therefore, performed with the host matrices reported in OQMD [75] and Materials Project databases [59]. However, some high-performance thermoelectrics (like certain polymorphs) are actually *metastable* (slightly above the hull distance) which can be synthesized through specific "non-equilibrium" processing routes, such as, High-Pressure Torsion (HPT) or Spark Plasma Sintering (SPS), to stabilize phases that would otherwise decompose under slow cooling.

Given that strict filtering removes them; and loose filtering includes too much junk, finding the right "Distance to Hull or $\Delta E_{hull}$" threshold is a major challenge. By convention, researchers use a simple rule of thumb to guess if a material can be synthesized: they look for an energy level roughly one to four times that of room temperature $k_B T$ —specifically between 25 and 100 meV.atom$^{-1}$ [75]. This "soft criterion" is based on the rough assumption that heat (entropy) at normal temperatures can compensate for that amount of instability. These conservative estimates are intentionally low to ensure a high probability of success. However, data from approximately 30,000 inorganic materials in the Materials Project reveals that this "one-size-fits-all" approach is flawed [76]. In reality, the energy limit for synthesis varies wildly depending on the type of material. For instance, most synthesized metastable oxides and nitrides fall within a much broader range—anywhere from 0.05 to 0.2 eV/atom [75, 76]—proving that some material classes can be much more unstable than others and still be successfully created. TeFeSb, for example, has a reported hull distance of 0.125 and 0.66 eV/atom in OQMD and Materials Project databases, respectively. However, this half-Heusler system has been synthesized experimentally and is one of the best performing TE materials in the hH family ($zT$ of ~1.4 for $Ta_{0.84}Ti_{0.16}FeSb$ at 970 K) [77]. Moreover, these databases don't essentially cover the entire possible combinatorial chemical spaces of different material classes. As we saw in Figure 2 (Section 1.a), for hH materials, adding host matrices from Materials Project (MP) barely improved the coverage of the possible combinatorial space. Therefore, a sole reliance on these databases deprives us from the opportunity for a *true* high-throughput screening and, thereby, reduces the probability of discovering promising TE compositions.

Determining the phase stability of a hypothetical compound from *ab initio* calculations requires screening of all possible phases for a given composition for the lowest enthalpies of formation. This task becomes prohibitively expensive for high-throughput *ab initio* calculations of vast compositional spaces especially without compromising the limited precision afforded by DFT in predicting the stability of the screened compositions [78, 79]. In this case too, ML can facilitate the phase prediction for a given chemical composition. In high-throughput screening, they can serve as a "fast filter" to identify promising candidates before committing to laboratory synthesis. The Matbench Discovery leaderboard [80] dubbed as the "Olympics" of materials stability prediction ranks several - MLIP based- models, such as -eSEN-30M-OAM [81], EquFlash [82], and Nequip-OAM-XL [83], based on their ability to act as a filter for materials discovery. Unlike standard benchmarks that just measure "how close the energy prediction is",

Matbench Discovery measures "how many stable materials the model actually finds." Besides, industry giants such as GNoME [84], CHGNet [85] and M3GNet [86] are also very popular. A strategic approach to reduce uncertainty would involve 'ensemble averaging' [87] for the regression of formation energy ($\Delta H_f$) or distance-to-hull ($\Delta E_{hull}$), and 'ensemble voting' [88] for categorical stability classification. A low standard deviation in averaged energy values, or high consensus among top-ranked models on the leaderboard, can serve as a robust *a priori* estimate of prediction reliability. However, since 0 K distance-to-hull is the current benchmark, it risks overlooking high-temperature stable phases crucial for thermoelectric materials. Next-generation 'fast filters' should incorporate Temperature-Dependent Effective Potential (TDEP) calculations [89, 90], leveraging universal MLIPs, such as CHGNet, to assess dynamic stability under realistic operating conditions (500–800 K). This approach effectively replaces traditional static hulls with a temperature-aware convex hull stability framework, reducing the risk of false negatives in the discovery of high-performance TE materials.

.

Contrary to metastable materials, a material may be computationally "stable" yet experimentally "unsynthesizable" due to dynamic chemical behaviors in high-temperature synthesis, such as elemental volatility and defect chemistry, not captured by standard DFT or MLIP methods. Therefore, in addition to using these models for phase prediction, combinatorial synthesis via. thin film material libraries (TFML) [91] can be used as a precursor for the most challenging step of the ML workflow: bulk synthesis. In the context of thermoelectric (TE) materials, where we often screen complex alloys or specific crystal structures (like half-heuslers), this technique allows "rapid mapping" of an entire chemical system in a single experiment (hundreds of different compositions on a single wafer). By using high-throughput X-ray Diffraction (XRD), one can quickly see where the desired phase forms and where it decomposes into secondary phases. Since, thin-film growth is a "quenched" process, this technique also allows for the discovery of metastable phases by "trapping" materials in high-energy states that might be difficult to reach via bulk melting. While bulk synthesis remains necessary to fully optimize lattice thermal conductivity and microstructure, the TF-to-bulk pipeline is a proven strategy for identifying viable 'stable phases' while minimizing the time-intensive trial-and-error traditionally associated with bulk TE discovery [92]. The limitations in transferability from thin film to bulk synthesis can further be minimized by selecting only the most "stable" compositions from the film library for expensive, time-consuming bulk synthesis (SPS/Arc-melting).

## 4. The final active learning loop

Summing up the strategies discussed in the previous sections, to bridge the gap between computational screening and experimental synthesis, we propose a self-consistent integrated active learning (AL) framework as shown in Figure 3. The HTS process begins with C2S (Composition-to-Structure) models to generate structural features (for structure-aware models) or *a priori* validate the structure (for structure-specific models) for candidates with hypothetical compositions. These candidates can be immediately mapped via Principal Component Analysis (PCA) to evaluate their position within the known chemical search space. At this step, we propose a phased expansion of the searchable chemical space by integrating PCA-derived distance metrics into Expected Improvement (EI) scores within a Bayesian Optimization (BO) framework [93]. Though the choice of acquisition function is not strict [94], Jang et al (2025) [95] have reported promising results for the experimental discovery of high entropy TE chalcogenides using EI. In our modified approach, candidates are prioritized based on their EI scores, weighted by their proximity to the training data distribution. This strategy ensures that the active learning loop initially targets high-confidence 'proximity' zones to validate model interpolation before systematically pushing toward under-sampled regions, or 'PCA boundaries.

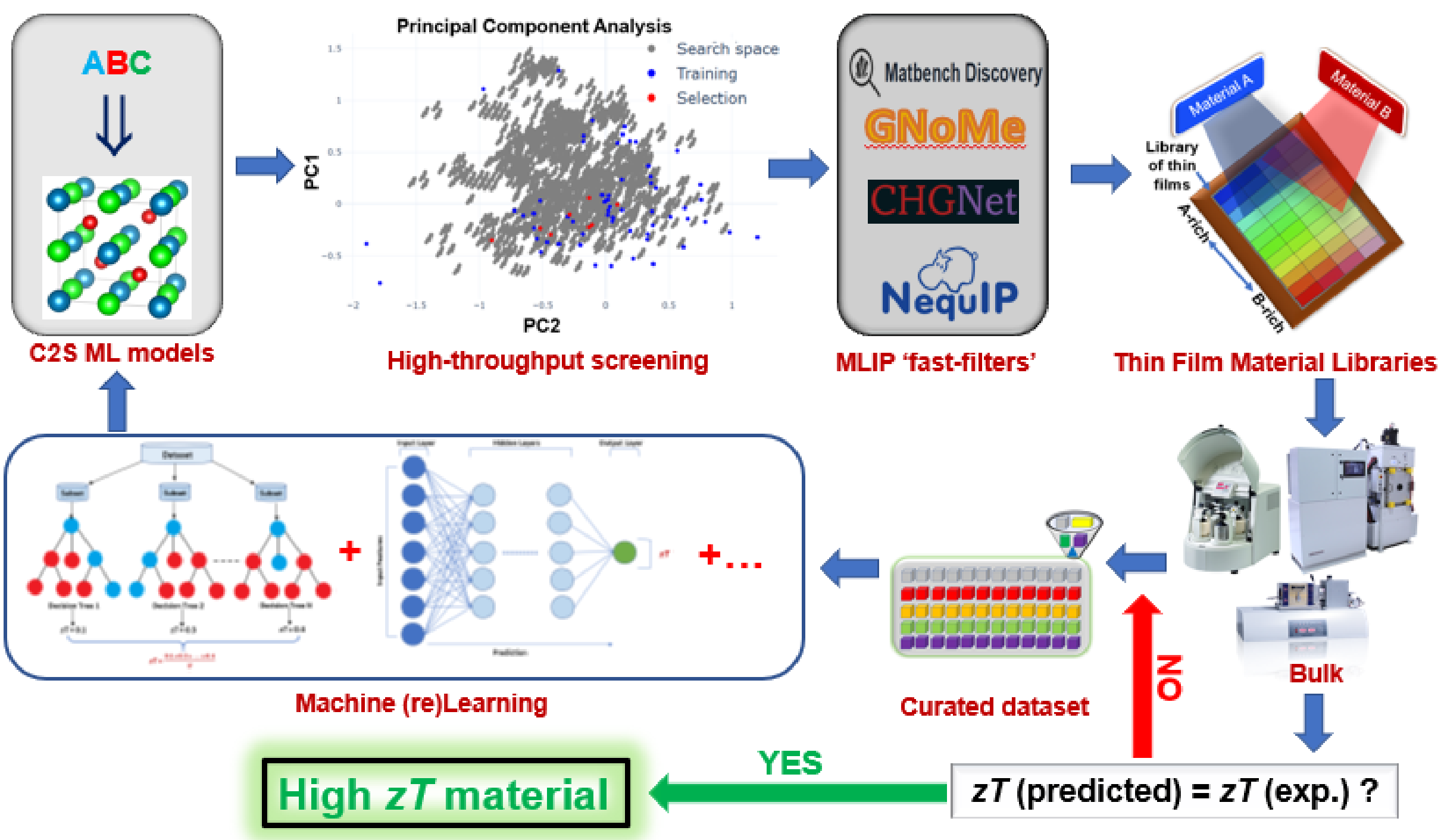

Figure 3: The active learning loop to bridge the gap between ML-predictions and experimental validations

These frontiers offer both maximum information gain—as no experimental data currently exists—and a high probability for discovering materials that can break known performance ceilings. This gradual expansion facilitates the optimal use of experimental resources while providing a rigorous framework to evaluate the limits of a model's generalizability across diverse chemical spaces. Selected candidates are then subjected to high-fidelity MLIP 'fast-filters'—including GNoMe, CHGNet, and NequIP—to ensure thermodynamic stability before proceeding to experimental validation. The synthesis stage utilizes combinatorial thin-film material libraries to rapidly explore compositional gradients, followed by characterization and potential scale-up to bulk forms. A critical feedback loop compares experimental $zT$ values against predictions; if a discrepancy exists, the results are funneled into a curated dataset for machine (re)learning. In contrast to incremental or transfer learning, batch re-training is the most rigorous weight update strategy because it optimizes the model against the entire cumulative data pool [84]. By maintaining a globally consistent optimization objective, it effectively eliminates model drift and catastrophic forgetting [96, 97]. This ensures that the model preserves the fundamental physical relationships of established chemical families while accurately integrating new experimental insights. In summary, the proposed AL framework allows systematic refinement of the predictive accuracy of ML models such that each iteration into novel chemical territory is backed by increased statistical and thermodynamic confidence.

## 5. Conclusion

The integration of machine learning into the discovery of thermoelectric (TE) materials represents a paradigm shift, yet the transition from computational prediction to successful **experimental realization** remains fraught with challenges. As this review has demonstrated, the primary obstacles—**the "small-data" problem, sampling bias, inadequate structure representation, and phase stability**—must be addressed to move beyond theoretical $zT$ optimization toward tangible experimental outcomes.

To enhance model generalizability, it is no longer sufficient to rely on standard randomized cross-validation, which often yields over-optimistic results. Instead, researchers must adopt **clustering and PCA-based validation strategies** that prioritize chemical and structural diversity. Furthermore, while "structure-agnostic" models offer speed, limiting ML to specific structural prototypes is essential for accurately capturing the transport properties of complex alloys and doped systems without relying on large structural feature spaces.

Perhaps the most significant constraint identified is **thermodynamic stability**. A high predicted $zT$ is of little value if the material cannot be synthesized. By utilizing ML "fast filters" like GNoME rather than relying on arbitrary "distance to Hull" criteria and public databases, researchers can better identify viable candidates from the exhaustive combinatorial space. The proposed synergy between **active learning** and **combinatorial thin-film synthesis**, offers a powerful roadmap; thin films can serve as a rapid, cost-effective precursor to bulk synthesis, and active learning can effectively enrich TE datasets by generating compositionally diverse data whilst improving model generalizability. Ultimately, a collaborative effort to build high-quality, diverse experimental datasets will be the cornerstone of a truly data-driven revolution in thermoelectric energy harvesting.

## Acknowledgements

We acknowledge the financial support from the Agence Nationale de la Recherche (ANR), France under the ANR-DFG project « CombiHeusler » (ANR-24-CE92-0053-01).

## Competing Interests

The authors declare no Competing Financial or Non-Financial Interests.